\documentclass[apj,tighten,iop,twocolumn]{emulateapj}

\newcommand{\mc}{\multicolumn}

\newcommand{\ltsimeq}{\raisebox{-0.6ex}{$\,\stackrel 
        {\raisebox{-.2ex}{$\textstyle <$}}{\sim}\,$}} 
\newcommand{\gtsimeq}{\raisebox{-0.6ex}{$\,\stackrel 
        {\raisebox{-.2ex}{$\textstyle >$}}{\sim}\,$}}

\newcommand{\mgii}{Mg\,{\sc ii}}

\newcommand{\lya}{Ly\,$\alpha$}
\newcommand{\lyb}{Ly\,$\beta$}

\newcommand{\nv}{N\,{\sc v}}
\newcommand{\hbeta}{H\,$\beta$}

\newcommand{\mbh}{$M_{\rm BH}$}

\newcommand{\myemail}{chris.willott@nrc.ca}

\def\co21{CO\,(2-1)}

\shorttitle{Eddington-limited accretion and the black hole mass function at redshift 6}
\shortauthors{Willott et al.}

\begin{document}


\title{Eddington-limited accretion and the black hole mass function at redshift 6}


\author{
Chris J. Willott\altaffilmark{1},
Loic Albert\altaffilmark{2},
Doris Arzoumanian\altaffilmark{3},
Jacqueline Bergeron\altaffilmark{4},
David Crampton\altaffilmark{1},
Philippe Delorme\altaffilmark{5},
John B. Hutchings\altaffilmark{1},
Alain Omont\altaffilmark{4},
C\'eline Reyl\'e\altaffilmark{6},
and David Schade\altaffilmark{1},
}

\altaffiltext{1}{Herzberg Institute of Astrophysics, National Research Council, 5071 West Saanich Rd, Victoria, BC V9E 2E7, Canada; \myemail}
\altaffiltext{2}{Canada-France-Hawaii Telescope Corporation, 65-1238 Mamalahoa Highway, Kamuela, HI96743, USA}
\altaffiltext{3}{CEA-Saclay, IRFU, SAp, 91191, Gif-sur-Yvette, France} 
\altaffiltext{4}{Institut d'Astrophysique de Paris, CNRS and Universit\'e Pierre et Marie Curie, 98bis Boulevard Arago, F-75014, Paris, France}
\altaffiltext{5}{School of Physics \& Astronomy, University of St Andrews, North Haugh, St Andrews KY16 9SS, UK}
\altaffiltext{6}{Institut Utinam, Observatoire de Besan\c{c}on, Universit\'e de Franche-Comt\'e, BP1615, 25010 Besan\c{c}on Cedex, France}

\begin{abstract}

We present discovery observations of a quasar in the Canada-France
High-$z$ Quasar Survey (CFHQS) at redshift $z=6.44$.  We also use
near-IR spectroscopy of nine CFHQS quasars at $z \sim 6$ to determine
black hole masses. These are compared with similar estimates for more
luminous Sloan Digital Sky Survey (SDSS) quasars to investigate the
relationship between black hole mass and quasar luminosity. We find a
strong correlation between \mgii\ FWHM and UV luminosity and that most
quasars at this early epoch are accreting close to the Eddington
limit. Thus these quasars appear to be in an early stage of their life
cycle where they are building up their black hole mass
exponentially. Combining these results with the quasar luminosity
function, we derive the black hole mass function at $z=6$. Our black
hole mass function is $\sim 10^{4}$ times lower than at $z=0$ and
substantially below estimates from previous studies. The main
uncertainties which could increase the black hole mass function are a
larger population of obscured quasars at high-redshift than is
observed at low-redshift and/or a low quasar duty cycle at $z=6$. In
comparison, the global stellar mass function is only $\sim 10^{2}$
times lower at $z=6$ than at $z=0$. The difference between the black
hole and stellar mass function evolution is due to either rapid early
star formation which is not limited by radiation pressure as is the
case for black hole growth or inefficient black hole seeding. Our work
predicts that the black hole mass -- stellar mass relation for a
volume-limited sample of galaxies declines rapidly at very high
redshift. This is in contrast to the observed increase at $4<z<6$ from
the local relation if one just studies the most massive black holes.
\end{abstract}


\section{Introduction}

Quasars are very luminous objects residing at the centers of galaxies
believed to be powered by accretion of matter onto a supermassive
black hole. The high luminosity makes them identifiable out to very
high redshift where they can be used to study black hole growth,
galaxy evolution and the intergalactic medium. It has been discovered
that most nearby, massive galaxies harbor central supermassive black
holes and that the mass of the black hole scales with galaxy
properties such as bulge luminosity (Magorrian et al. 1998) and
velocity dispersion (Ferrarese \& Merritt 2000; Gebhardt et
al. 2000). These observations suggest that black holes play an
important role in galaxy evolution, likely due to feedback which heats
and expels gas which would otherwise form stars in the galaxy (Silk \&
Rees 1998; Di Matteo et al. 2005, Hopkins et al. 2006a; Croton et
al. 2006). The physical details of how the energy emitted by the
active galactic nucleus (AGN) couples to the gas are still poorly
understood.

The relic black hole mass function at low redshift is the result of
all the black hole growth via accretion and merging over cosmic
time. There is good agreement between the number density of black
holes at low redshift and the observed accretion as inferred by
studies of AGN luminosity functions if black holes accrete with a
mass-to-energy conversion efficiency in the range $0.06 \ltsimeq \epsilon \ltsimeq 0.10$ (Yu \&
Tremaine 2002; Marconi et al. 2004; Hopkins et al. 2007; Shankar et
al. 2009). Such studies also agree with luminosity function
observations at various redshifts (e.g. Ueda et al. 2003) that there is {\it downsizing} in
the AGN population, that the more massive black holes were
built up more rapidly at high-redshift than the less massive black
holes which accreted a greater fraction of their mass at lower redshift.

Reverberation mapping of broad-lined AGNs at low-redshift has shown
that a relationship exists between the distance of the line-emitting
gas from the central ionizing sources, $R$, and the optical/UV
luminosity, $L$ (Kaspi et al. 2000; Bentz et al. 2009). This has the
form $R \propto L^{0.5}$, as expected based on simple ionization
models. The consequence of this relationship is that measurement of
the velocity of the line-emitting gas, e.g. via its Doppler-broadened
linewidth, and the luminosity are sufficient to determine the
gravitational mass of the central black hole (Wandel et
al. 1999). Such measurements can be made with single epoch, moderate
signal-to-noise (S/N) spectra out to very high redshifts. This method
appears to give a fairly low scatter (between 0.2 to 0.3 dex for the
\hbeta\ and \mgii\ lines; Kollmeier et al. 2006; Fine et al. 2008;
Shen et al. 2008; Steinhardt \& Elvis 2010), comparable with that of
reverberation mapping. For a review of the reliability and
accuracy of this method we refer the reader to Peterson (2010).

Using this method it has been possible to measure black hole masses
for SDSS quasars at $z=6$ (Willott et al. 2003; Jiang et al. 2007;
Kurk et al. 2007; Kurk et al. 2009). Most of these are from the main
SDSS sample of Fan et al. (2006) which contains very luminous
quasars. These results show that the most luminous $z=6$ quasars
contain black hole masses $M_{\rm BH} > 10^9 M_\odot$ accreting at
close to the Eddington limit. This is not too surprising because if
they were to accrete at substantially below the Eddington limit, the
required black hole masses would be $M_{\rm BH} > 10^{10} M_\odot$ and
such black holes are rare even at more moderate redshifts (McLure \&
Dunlop 2004; Vestergaard \& Osmer 2009). If there are sub-Eddington
accreting black holes at $z=6$ they would be found in the lower
luminosity quasar samples such as the SDSS deep stripe (Jiang et
al. 2009) and Canada-France High-$z$ Quasar Survey (CFHQS; Willott et
al. 2010). Two of the SDSS deep stripe quasars have black hole mass
measurements and both are found to have lower black hole masses than
the SDSS main quasars and are accreting at approximately the Eddington
limit (Kurk et al. 2007; 2009). In order to increase the numbers of
low luminosity quasars with black hole mass measurements and determine
if any of them are powered by high mass black holes accreting at low
Eddington ratios, we are carrying out a near-IR spectroscopy program
on all $z>6$ CFHQS quasars.

The existence of $M_{\rm BH} > 10^9 M_\odot$ black holes at redshift 6
has led to much theoretical work to explain how such objects can be
built up in the $<1$\, Gyr of cosmic time available. Standard
Eddington-limited accretion has an e-folding time (the Salpeter
time-scale) of $4.5 \times 10^7$\,yr assuming efficiency
$\epsilon=0.1$. This led some people to propose that episodes of
super-Eddington accretion at very high redshift were necessary
(Volonteri \& Rees 2005; Kawakatu \& Wada 2009). Another possibility
is higher efficiency which lowers the Salpeter time-scale and may be
expected for rapidly spinning black holes (Shapiro 2005). Other
studies have shown that Eddington-limited accretion can account for
these black holes, but only if very massive seed black holes or
multiple stellar seeds are invoked (Yoo \& Miralda-Escud\'e 2004;
Sijacki et al. 2009). Mergers are also an important part of the growth
of the most massive black holes at high redshift (Volonteri \& Rees
2006; Li et al. 2007).

An important prediction of galaxy evolution models is the evolution of
the black hole mass -- stellar mass relation. There have been many
attempts at observational determinations of this ratio, however
selection effects are critical. Studies involving AGN tend to show
higher than local ratios of $M_{\rm BH}/M_{\rm stellar}$ at all
redshifts; $z<1$ (Mathur \& Grupe 2005; Woo et al. 2008), $1<z<4$ (Peng et
al. 2006; McLure et al. 2006; Merloni et
al. 2010); $z>4$ (Walter et al. 2004; Riechers et al. 2008; Wang et al. 2010), whereas studies based on $z\approx 2$ starburst
galaxies show lower ratios (Borys et al. 2005; Alexander et
al. 2008). Other approaches consider the global evolution of the
stellar mass function and black hole mass function using various
constraints such as the observed quasar luminosity function, the relic
black hole mass function, the X-ray background or results from
theoretical simulations (Hopkins et al. 2006b; Di Matteo et al. 2008;
Somerville 2009; Shankar et al. 2009). Such studies have shown fairly
little evolution in this ratio, at least out to $z=2$ where they are
most strongly constrained by data.

In this paper we present new data on CFHQS quasars designed to address
the growth of black holes at early times. In section 2 we present the
discovery of the most distant known quasar at $z=6.44$. Near-IR
spectroscopy of nine CFHQS quasars is presented in Section 3. In
Section 4 we use these data to derive black hole masses, investigate
correlations between FWHM, luminosity and black hole mass and
determine the Eddington ratio distribution and its
implications. Section 5 combines this work with the $z=6$ quasar
luminosity function of Willott et al. (2010) to determine the black
hole mass function and compare its evolution with that of the global
stellar mass function. In the appendix we show data on four quasars
for which no black hole mass measurement was possible and discuss why.

All optical and near-IR magnitudes in this paper are on the AB
system. Cosmological parameters of $H_0=70~ {\rm km~s^{-1}~Mpc^{-1}}$,
$\Omega_{\mathrm M}=0.28$ and $\Omega_\Lambda=0.72$ (Komatsu et
al. 2009) are assumed throughout.


\begin{deluxetable*}{c l c c c c c c}
\tablewidth{500pt}
\tablecolumns{8}
\vspace{0.4cm}
\tablecaption{\label{tab:photom} New quasar position  and photometry} 
\tablehead{ Quasar    & RA and DEC (J2000.0)     &  $i'$ mag        &    $z'$ mag       &      $J$ mag      &  $i'-z'$        &  $z'-J$ & $M_{1450}$     }
\startdata
CFHQS\,J021013-045620 &  02:10:13.19 -04:56:20.9 & $>25.74^{\rm a}$ &  $22.67 \pm 0.05$ & $22.28 \pm 0.27$  & $>3.07$ & $0.39 \pm 0.28$  & $-24.28$ 
\enddata
\tablecomments{All magnitudes are on the AB system.}
\end{deluxetable*}

\begin{figure}
\resizebox{0.48\textwidth}{!}{\includegraphics{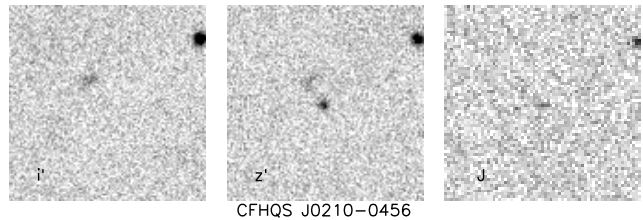}}
\caption{Images in the $i'$, $z'$ and $J$ filters centered on CFHQS \,J0210-0456. Each image covers $20'' \times 20''$. The images are oriented with north up and east to the left. 
\label{fig:cutouts}
}
\end{figure}

\begin{figure}[b]
\resizebox{0.48\textwidth}{!}{\includegraphics{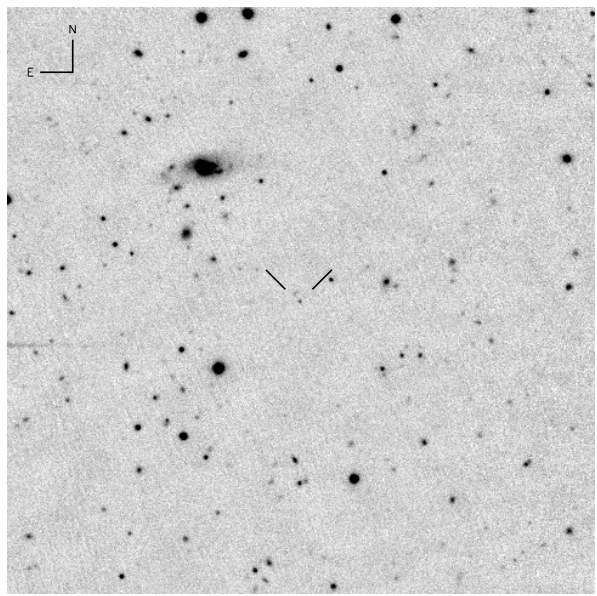}}
\vspace{-1.0cm}
\caption{ $z'$-band finding chart for CFHQS \,J0210-0456. The field of view is $3' \times 3'$.
\label{fig:finders}
}
\end{figure}

\begin{figure*}
\hspace{0.5cm}
\resizebox{0.93\textwidth}{!}{\includegraphics{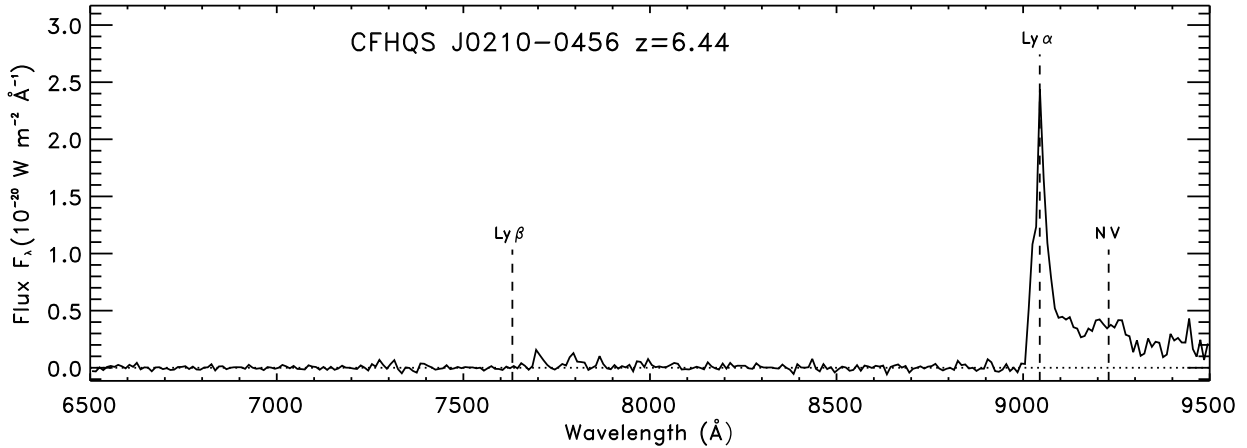}}
\caption{Optical spectrum of the newly discovered quasar. The
locations of \lya, \lyb\ and \nv\ for a redshift of $z=6.44$ are marked with dashed
lines.  The spectrum is binned in 10\,\AA\ pixels. 
\label{fig:spec}
}
\end{figure*}

\section{Discovery of a new CFHQS quasar at $z=6.44$}

In previous papers, we have presented imaging and optical spectroscopy
of 19 CFHQS quasars. Since these publications, we have discovered one
more CFHQS quasar. In this section we present the observations and
features of this quasar. The quasar was discovered by its red $i'-z'$
color in imaging in the CFHT Legacy Survey Wide W1 field\footnotemark\
\footnotetext{http://www.cfht.hawaii.edu/Science/CFHTLS}. Follow-up
near-IR imaging observations at the 3.6m ESO New Technology Telescope
with the SOFI instrument showed that it has a relatively blue $z'-J$
color and is therefore more likely to be a high-redshift quasar than a
T dwarf (see Figure 1 of Willott et al. 2009). Figure
\ref{fig:cutouts} shows small regions centred on this object in the
$i'$, $z'$ and $J$ filters. The object is not detected at $i'$ and
only weakly detected at $J$. A larger $z'$-band field is shown in
Figure \ref{fig:finders} which can be used as a finding
chart. Photometry and the position of this quasar, named
CFHQS\,J021013-045620, are given in Table \ref{tab:photom} (note that from
here on we use the abbreviated version of the name).

Spectroscopy of the quasar was obtained on 2009 November 12 and 2009
December 17 with the GMOS spectrograph at the Gemini-South
telescope. The R400 grating was used with a 1 arcsec slit to give a
resolving power of 1300. The seeing was 0.8 arcsec and the sky
transparency was photometric. The total integration time was 3.5
hours.  The nod-and-shuffle mode was employed to enable very accurate
sky subtraction.  The data reduction method is as described in Willott
et al. (2007).

Figure \ref{fig:spec} presents the reduced optical spectrum of
CFHQS\,J0210-0456. The most obvious feature in the spectrum is the
strong, narrow, asymmetric \lya\ emission line. The peak of this
emission line is at 9045 \AA\ which is equal to a redshift of
$z=6.44$.  Accounting for some absorption on the blue wing of the
line, the estimated intrinsic FWHM of the line is $1300 ~{\rm
  km\,s^{-1}}$. The redshift of the peak does not change appreciably
when making such an absorption correction (peak shifts to 9040
\AA). This spectrum shows several other characteristics of $z>6$
quasars. There is a sharp break in the continuum level across the
\lya\ line due to foreground neutral hydrogen absorption. The
continuum flux at observed-frame 7800 \AA\ is clearly visible which
corresponds to light allowed to pass through the IGM at $z\approx
5.4$. Shortward of the expected location of \lyb\ there is complete
absorption again. There is a possible \nv\ broad emission line which
is marked on Figure \ref{fig:spec}.

In the following section we will describe near-IR spectroscopy of a
sample of CFHQS quasars. One of the quasars observed is
CFHQS\,J0210-0456. Although this quasar is extremely faint in the
near-IR, a 4 hours integration with NIRI on Gemini-North allowed the
detection of the \mgii\ $\lambda 2799$ emission line (Figure
\ref{fig:nirispec}). The best-fit \mgii\ redshift is $z=6.438 \pm
0.004$. This is statistically identical to the redshift of the peak of
\lya. The closeness of the \lya\ peak and \mgii\ redshifts has been
noted before for many CFHQS quasars (Willott et al. 2010) and will be
investigated in a future paper. This redshift of $z=6.438$ is the
highest of any known quasar, just surpassing SDSS\,J1148+5251 at
$z=6.419$ (Fan et al. 2003; Walter et al. 2003) and CFHQS\,J2329-0301
at $z=6.417$ (Section \ref{nirspec}).

The accurate redshift for CFHQS\,J0210-0456 allows us to measure the
size of the ionized region in front of the quasar via the physical
extent of measurable flux on the blue side of \lya. This can give
important information on the ionization state of the IGM surrounding
the quasar (Cen \& Haiman 2000; Wyithe \& Loeb 2004; Bolton \&
Haehnelt 2007). The spectrum of CFHQS\,J0210-0456 reaches zero flux at
9005 \AA, which corresponds to a ionized near-zone size of just 1.7
proper Mpc. Note that using the wavelength at which the transmission
reaches 10\%, as used by Fan et al. (2006), rather than where the flux
reaches zero, gives almost exactly the same size due to the steepness
of the spectrum. Given this small size and steep drop, we do not bin
the spectrum in 20 \AA\ bins (as Fan et al. did) since then the
measured size would be very dependent upon the choice of bins.

This size of 1.7 Mpc is extremely small, much smaller than any other
known $z=6$ quasars with the exception of lineless or BAL quasars and
less than the typical size for SDSS quasars at $z>6.1$ of 5 Mpc
(Carilli et al. 2010). Note however that if these near-zone sizes are
set by the expansion of an ionization front from the quasar then they
are expected to scale with the quasar luminosity as $R \propto
L^{1/3}$. Since CFHQS\,J0210-0456 has a luminosity a factor of ten
lower than SDSS quasars, then the luminosity-scaled near-zone size is
consistent with that expected based on the near-zone sizes of the SDSS
$z>6.1$ quasars. Further exploration of the near-zone size and other
constraints on the ionization state of the IGM using this quasar will
be presented elsewhere.

\section{Near-infrared spectroscopy observations}
\label{nirspec}

We are attempting $K$-band spectroscopy of all CFHQS quasars with
redshifts $z>6$. We do not target the $z<6$ quasars because the
\mgii\ line center lies at wavelengths $<1.96 \mu$m and there is
substantial atmospheric absorption shortward of this wavelength which
could affect determining a reliable continuum and line width. Of the
20 quasars found in the CFHQS so far, 5 were not targeted because they
had $z<6$. Two $z>6$ quasars have not yet been targeted due to their
recent discovery. Therefore a total of 13 quasars have had near-IR
spectroscopy attempted.

One quasar, CFHQS\,J1509-1749, already has a published near-IR
spectrum from the GNIRS instrument at Gemini-South (Willott et
al. 2007). Near-infrared spectroscopy of the other 12 quasars was
obtained using the NIRI instrument on the Gemini-North telescope. The
$K$-band grism covers the wavelength range $1.87 - 2.59\,\mu$m at a
resolution of $R=520$ (570 ${\rm km\,s^{-1}}$) with the 0.75 arcsec
slit. The seeing was generally good -- in the range 0.5 to 0.7
arcsec. The observations of each target were split into 300\,second
frames nodded along the slit to enable good sky subtraction and bad
pixel rejection. All frames were checked for the fixed pattern noise
problem that randomly affects some NIRI exposures. Any pattern noise 
affected frames or frames taken in variable sky conditions were 
discarded. Total usable integration times ranged from 1.3 to 4
hours with an average of 2.8 hours.

Data reduction followed standard near-IR longslit procedures which
will be briefly described here. Flat-field frames obtained from
continuum lamp exposures were combined to make a flat-field image used
to flatten the individual 300\,second frames. First-pass sky
subtraction was performed by determining the 2D sky spectrum from
neighboring frames obtained within a time-span of
$\pm\,450$\,seconds. Residual sky emission was removed by fitting a
spline along the column of each spectral row. The spatial shifts
between the individual frames were determined via measurement of the
quasar location to enable all the frames to be combined into a single
2D spectrum. The combination process averaged the pixel fluxes using
positive and negative sigma clipping and bad pixel mask rejection.
Wavelength calibration was performed using a 4th order polynomial fit
to an argon lamp spectrum. The residuals of this fit about the known
wavelengths were around 0.7\AA. Relative flux calibration and
atmospheric absorption corrections were made using spectra of early
A-type stars observed at similar airmass. Absolute flux calibration
was performed using broad band photometry. The uncertainty as a
function of wavelength was determined by measuring an iterative
sigma-clipped rms along the clean parts of each column (spatial
direction) in the 2D spectrum.

\begin{figure*}
\resizebox{1.00\textwidth}{!}{\includegraphics{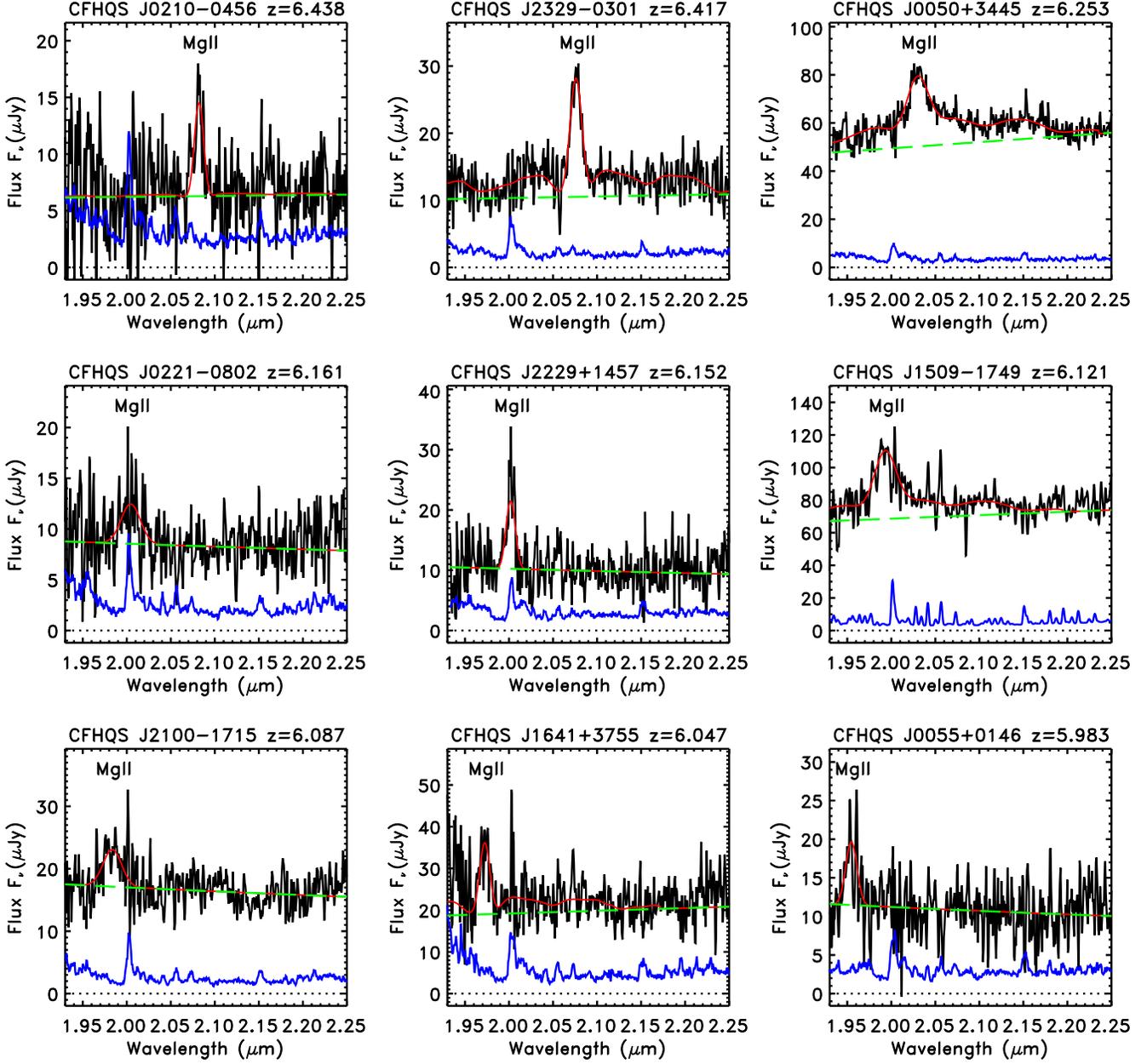}}
\caption{$K$-band spectra of CFHQS quasars (black line). The fitted
  model of power law continuum, broadened Fe template and broadened
  \mgii\ doublet is shown as a red line. The power law continuum only
  is shown as a green dashed line. At the bottom of the plot, the blue
  line is the $1\sigma$ noise spectrum. The location of the
  \mgii\ broad emission line is labeled. Note how narrow most of the 
  \mgii\ lines are.
\label{fig:nirispec}
}
\end{figure*}

Fig\,\ref{fig:nirispec} shows the resulting spectra of 9 of the 13
observed quasars and the $1\,\sigma$ uncertainty.  The S/N per pixel
in the continuum at clean parts of the spectra ranges from 3 to 20
depending upon the source flux-density, total integration time and
observing conditions. For the other 4 quasars the spectra are not
suitable for the measurement of the \mgii\ line due to insufficient
S/N. These observations are discussed in the appendix. As explained
there, we do not believe the exclusion of these quasars from the
following analysis causes any biases. Although at first sight there
appear to be \mgii\ associated absorption lines for some quasars, none
are detected at high significance. Most apparent absorption lines are
at wavelengths of night sky emission lines where the noise is
higher. These will not strongly affect the fitted \mgii\ emission lines
because such data points have low weight in the fit.

The spectra consist of 3 possible components: a power law continuum, a
Doppler-broadened blend of many iron emission lines and the
\mgii\ 2799\AA\ emission line doublet. A simultaneous fit to these
three components was carried out. The iron line emission template is
from McLure \& Dunlop (2004) and covers the spectral range 2100 --
3085 \AA. Therefore only data in this rest-frame wavelength range was
included in the fit. There are six free parameters in the fit: (i) the
continuum normalization; (ii) the continuum power law slope; (iii) the
redshift of the iron and \mgii\ broad emission lines (assumed equal);
(iv) the normalization of the iron template; (v) the peak height of
the Gaussian \mgii\ emission line; (vi) the FWHM of the
\mgii\ doublet.

The best-fit was determined by minimizing the $\chi^2$ using an amoeba
algorithm (Press et al. 1992). The best-fit continuum and total
spectra are plotted on top of the data in Figure \ref{fig:nirispec}. In
all cases, the fits provide a good description of the data. Bootstrap
resampling was used to derive uncertainties on the fitted
parameters. 500 randomly-perturbed realizations of each spectrum were
generated and fit by $\chi^2$ minimization.

\begin{table*}
\caption{Measured and derived parameters of CFHQS quasars}
\vspace{-0.5cm}
\begin{center}
\begin{tabular}{ccccccc}
\colrule\colrule
\mc{1}{c}{Quasar} &\mc{1}{c}{\mgii\ redshift} &\mc{1}{c}{$M_{1450}$}  &\mc{1}{c}{FWHM (\mgii)}  &\mc{1}{c}{$L_{3000}$} &\mc{1}{c}{\mbh }    &\mc{1}{c}{$\lambda$} \\
\mc{1}{c}{ }      &\mc{1}{c}{$z\,_{\rm Mg\,{II}}$}&\mc{1}{c}{ }        &\mc{1}{c}{(km\,s$^{-1}$)} &\mc{1}{c}{($10^{37}$\,W)}           &\mc{1}{c}{($M_\odot$)} &\mc{1}{c}{($\equiv L_{\rm Bol}/L_{\rm Edd}$)}              \\
\colrule
CFHQS\,J0210-0456 & $6.438 \pm 0.004 $ & $-24.28$ & $1300\pm 350$  & $42\pm 5$   & $(8.0^{+5.5}_{-4.0})\times 10^{7} $  & $2.4$   \\
CFHQS\,J2329-0301 & $6.417 \pm 0.002 $ & $-25.00$ & $2020\pm 110$  & $71\pm 8$   & $(2.5^{+0.4}_{-0.4})\times 10^{8} $  & $1.3$  \\
CFHQS\,J0050+3445 & $6.253 \pm 0.003 $ & $-26.62$ & $4360\pm 270$  & $342\pm 34$ & $(2.6^{+0.5}_{-0.4})\times 10^{9} $  & $0.62$   \\
CFHQS\,J0221-0802 & $6.161 \pm 0.014 $ & $-24.45$ & $3680\pm 1500$ & $50\pm 5$   & $(7.0^{+7.5}_{-4.7})\times 10^{8} $  & $0.33$   \\
CFHQS\,J2229+1457 & $6.152 \pm 0.003 $ & $-24.52$ & $1440\pm 330$  & $60\pm 6$   & $(1.2^{+0.7}_{-0.5})\times 10^{8} $  & $2.4$   \\
CFHQS\,J1509-1749 & $6.121 \pm 0.002 $ & $-26.78$ & $4420\pm 130$  & $440\pm 44$ & $(3.0^{+0.3}_{-0.3})\times 10^{9} $  & $0.68$   \\
CFHQS\,J2100-1715 & $6.087 \pm 0.005 $ & $-25.03$ & $3610\pm 420$  & $98\pm 10$  & $(9.4^{+2.9}_{-2.5})\times 10^{8} $  & $0.49$   \\
CFHQS\,J1641+3755 & $6.047 \pm 0.003 $ & $-25.19$ & $1740\pm 190$  & $120\pm 13$ & $(2.4^{+1.0}_{-0.8})\times 10^{8} $  & $2.3$   \\
CFHQS\,J0055+0146 & $5.983 \pm 0.004 $ & $-24.53$ & $2040\pm 280$  & $63\pm 7$   & $(2.4^{+0.9}_{-0.7})\times 10^{8} $  & $1.2$   \\
\colrule
\end{tabular}
\end{center}
{\sc Notes.}---
\mgii\ redshifts are based on the wavelength of the fitted \mgii\ doublet and in all cases supersede previously published values which were based on \lya\ in the optical spectra. 
\label{tab:param}
\end{table*}

 Table\,\ref{tab:param} presents the best-fit \mgii\ redshift, FWHM
 velocities and their uncertainties for the nine CFHQS quasars. The broad
 line FWHM values in this table were determined by deconvolving the
 instrumental resolution (570 ${\rm km\,s^{-1}}$ for the NIRI spectra
 and 330 ${\rm km\,s^{-1}}$ for the GNIRS spectrum). Note that in all
 cases the deconvolved FWHM is greater than twice the instrumental
 FWHM meaning that the observations were well-resolved. The redshift
 uncertainty includes 0.002 added in quadrature due to the absolute
 wavelength calibration uncertainty of the spectroscopy. It does not
 include any extra uncertainty due to the fact the line shape may not
 be a perfect gaussian due to asymmetric emission or absorption
 lines.

\section{Black hole masses in $z\sim 6$ quasars}
\label{bh}

\subsection{Black hole mass determinations}
\label{bhdeterm}

The successful measurement of the \mgii\ emission line widths in these
quasars allows us to determine black hole masses by the virial
method. As described in Section 1, despite several potential sources
of uncertainty, this method appears to give black hole masses accurate
to $\approx 0.3$ dex (Shen et al. 2008; Steinhardt \& Elvis 2010) if
high quality line and continuum measurements are available. We use the
relationship determined by Vestergaard \& Osmer (2009) to calculate
black hole masses from the \mgii\ line width, FWHM\,(\mgii) and the
luminosity at rest-frame 3000\,\AA, $L_{3000}$\\

\begin{math}
\log M_{\rm BH}= 6.86+2\log \frac{\rm FWHM\,(Mg\,II)}{1000\,{\rm km\,s}^{-1}}+0.5\log \frac{L_{3000}}{10^{37}\,{\rm W}}.
\end{math}\\

\begin{figure*}
\resizebox{1.0\textwidth}{!}{\includegraphics{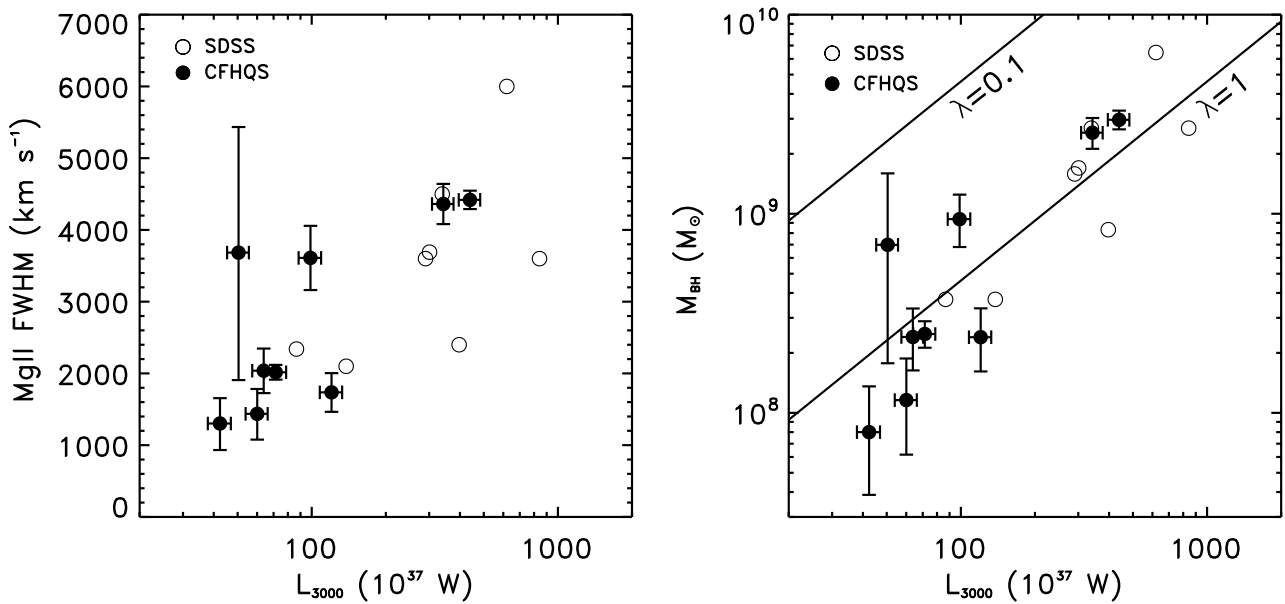}}
\caption{{\it Left:} \mgii\ FWHM against luminosity $L_{3000}$ for
CFHQS and SDSS $z\approx 6$ quasars. Error bars are only plotted for
the CFHQS quasars. There is a strong positive correlation between
the two independently-measured quantities. {\it Right:} Virial black
hole mass against luminosity for the same quasars. Also plotted are lines corresponding to Eddington ratios of $0.1$ and $1$. Quasars from both samples are accreting at close to the Eddington luminosity.
\label{fig:bhmassfwhmlum}
}
\end{figure*}

Values of $L_{3000}$ are derived from the fitted continuum
normalization and their uncertainties include 10\% added in quadrature
to account for the absolute flux calibration uncertainty. No
additional uncertainty due to variability was included. The black
hole mass uncertainties were determined by inserting the extreme
values of the luminosity and linewidth, based on their rms
uncertainties, in the above equation (rather than adding errors in
quadrature). No additional uncertainty was added to account for the
fact that the \mgii\ virial relation for black hole masses has its own
scatter of about 0.3 dex. The derived values for $L_{3000}$ and
$M_{\rm BH}$ are given in Table\,\ref{tab:param}.

As noted in Section \ref{nirspec}, the spectra have differing S/N due
to the fact that some of these quasars are very faint. Therefore we
have to consider whether the quasars with low S/N might have their
line width measurements biased high or low due to this. Shen et
al. (2008) studied virial black hole mass determinations from SDSS DR5
quasar spectroscopy. Their S/N distribution is similar to that of the
CFHQS spectra. They found that there was no dependence of the FWHM
distribution on the S/N of the spectra. Denney et al. (2009) found a
systematic shift towards lower measured black hole masses of $\sim
0.05$ dex as the S/N per pixel was artificially decreased from 20 to
5. Given that most of our spectra have S/N per pixel $>5$, we do not
consider this bias to be important for our study (0.05 dex is
substantially less than the \mbh\ uncertainty due to propagating the
measured line width uncertainties for the quasars with low S/N).  .

The CFHQS $z>6$ quasars are mostly of moderate optical luminosity. In
order to sample a broad range of luminosity at a fixed redshift we
also include in our analysis the published results of near-IR
\mgii\ spectroscopy of SDSS $z\approx 6$ quasars. The quasars used are
SDSS\,J1148+5251 $z=6.42$ (Willott et al. 2003), SDSS J1623+3112
$z=6.25$ (Jiang et al. 2007), SDSS J0005-0006 $z=5.85$, SDSS
J0836+0054 $z=5.81$, SDSS J1030+0524 $z=6.31$, SDSS J1306+0356
$z=6.02$, SDSS J1411+1217 $z=5.90$ (Kurk et al. 2007) and
SDSS\,J0303-0019 $z=6.08$ (Kurk et al. 2009). We take the values of
$L_{3000}$ and \mgii\ FWHM from these papers and use the above
equation to calculate the black hole mass in a consistent manner.
Even though this is not a complete sample of SDSS $z\approx 6$
quasars, the target selection was mostly done on the basis of redshift
and observability, so we do not expect any selection effects from the
incompleteness.  Thus our total sample consists of nine CFHQS quasars
and eight SDSS quasars which span a factor of 20 in luminosity.

In Figure \ref{fig:bhmassfwhmlum} we plot the \mgii\ FWHM and $M_{\rm
  BH}$ against $L_{3000}$ for these 17 quasars. It is immediately
apparent that there is a strong positive correlation between FWHM and
$L_{3000}$. This holds for both quasar surveys: the two high
luminosity CFHQS quasars have relatively high FWHM and the two
moderate luminosity SDSS quasars from the SDSS deep stripe (Jiang et
al. 2008) have low FWHM (as previously discussed by Kurk et
al. 2009). This strong correlation is particularly interesting because
it is not seen at lower redshifts (Fine et al. 2008; Shen et al. 2008)
where there is essentially no change in FWHM with luminosity. The fact
that there was no correlation of FWHM with luminosity could cast doubt
on the reliability of the virial black hole mass estimation method
since correlations between $M_{\rm BH}$ and luminosity are driven
by the fact that luminosity enters the equation used to estimate
$M_{\rm BH}$. The reason that, at lower redshifts, there is no
correlation between FWHM and luminosity is because the quasars are
accreting at a very wide range of Eddington ratios (Shen et
al. 2008). As we are about to show, this is not the case at $z=6$.

The Eddington luminosity is defined as the maximum luminosity attainable due to
outward radiation pressure acting on infalling material and is
$L_{\rm Edd} = 1.3\times10^{31}~ (M_{\rm BH}/M_\odot)$ W. The
observed monochromatic luminosity $L_{3000}$ is only a fraction of the
total electromagnetic luminosity coming from the quasar, so a
bolometric correction is applied to calculate the bolometric
luminosity, $L_{\rm Bol}$, from $L_{3000}$. We use a bolometric
correction factor of 6.0, consistent with previous work (Richards et
al. 2006; Jiang et al. 2006). The ratio of the bolometric luminosity
to the Eddington luminosity for a given black hole mass is often
called the Eddington ratio and is defined as $\lambda \equiv L_{\rm
  Bol}/L_{\rm Edd}$.

The right panel of Figure \ref{fig:bhmassfwhmlum} plots $M_{\rm BH}$
against $L_{3000}$. There is a strong positive correlation showing
that more luminous quasars have greater black hole masses. The black
hole masses in these $z=6$ quasars range from just under $10^8
M_\odot$ up to almost $10^{10} M_\odot$. Also shown are lines for
Eddington ratios of $\lambda=0.1$ and $\lambda=1$. The $z=6$
quasars are tightly clustered around the $\lambda=1$ line showing that
they are almost all accreting at the Eddington limit. There are no
quasars with $\lambda=0.1$ which are common at lower redshifts. The
distribution of $\lambda$ will be discussed further in the following
section. Note that some quasars appear to exceed the Eddington limit,
although only by up to a factor of 3.

Figure \ref{fig:bhmassfwhmlum} highlights why the observation of low
luminosity quasars at $z=6$ is so important. In the local universe
(Shankar et al. 2009) and at the peak of quasar activity at $1<z<3$
(McLure \& Dunlop 2004; Shen et al. 2008; Vestergaard \& Osmer 2009)
there appears to be a maximum black hole mass of $10^{10} M_\odot$,
likely due to feedback effects on black hole growth in the most
massive galaxies. The high luminosities of most $z\approx 6$ SDSS
quasars mean that they cannot be accreting at $\lambda=0.1$
because this would necessitate $M_{\rm BH} > 10^{10} M_\odot$. It is
the CFHQS quasars, with more moderate luminosities, that could be
powered either by $\sim 10^{9} M_\odot$ black holes with $\lambda=0.1$ or
$\sim 10^{8} M_\odot$ black holes with $\lambda=1$. Our results show that
the latter is true.

\subsection{The Eddington ratio distribution}
\label{eddratio}

The distribution of Eddington ratios, $\lambda$, in quasars is an
important constraint on models of quasar activity and black hole
growth. Studies up to $z=4$ show that the $\lambda$ distribution in
luminosity and redshift bins is usually a lognormal which shifts to
higher $\lambda$ and narrows at higher luminosities (Kollmeier et
al. 2006; Shen et al. 2008). The narrowing of the distribution at high
luminosity is likely related to the cutoff in the black hole mass
function at $\sim 10^{10} M_\odot$. Shen et al. (2008) found that the
most luminous quasars ($L_{\rm Bol}>10^{40}$ W) at $2<z<3$ have a
typical $\lambda=0.25$ and dispersion of 0.23 dex. Some authors have
discussed the Eddington ratio distribution as a function of black hole
mass (Kollmeier et al. 2006; Netzer et al. 2007) since this relates
more closely to model predictions. However, the selection effects
imposed by flux-limited quasar samples coupled with the correlations
of luminosity, \mbh\ and $\lambda$ have caused disagreement between
the results of these studies.

\begin{figure}
\resizebox{0.50\textwidth}{!}{\includegraphics{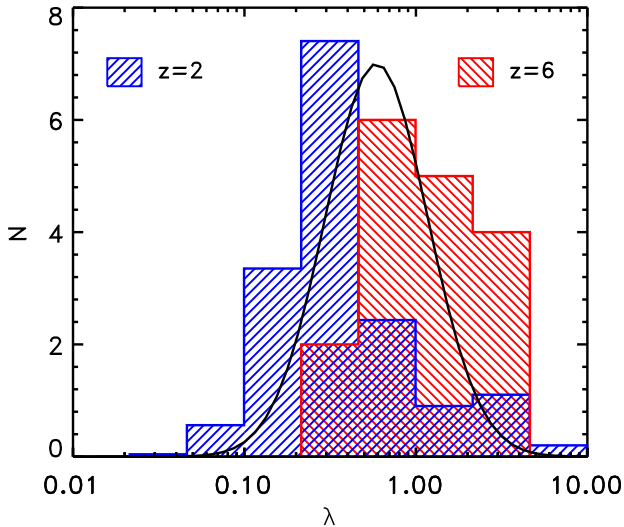}}
\caption{Observed distributions of the Eddington ratio for the
  luminosity-matched $z=2$ and $z=6$ quasar samples are shown as hatched
  histograms. The two distributions are clearly different showing that
  a much higher fraction of $z=6$ quasars are accreting at or just above
  the Eddington limit. The solid black line shows the intrinsic $\lambda$ distribution for $z=6$ black holes after accounting for the fact that high-$\lambda$ black holes are more likely to be found in flux-limited quasar samples than those with low values of $\lambda$.
\label{fig:eddfrac}
}
\end{figure}

The black hole masses of the 17 $z=6$ quasars allow us to make the
first determination of the $\lambda$ distribution at such a high
redshift. We do not have sufficient quasars to determine the
distribution as a function of luminosity, however inspection of Figure
\ref{fig:bhmassfwhmlum} suggests no luminosity-dependence. In Figure
\ref{fig:eddfrac} we show the $\lambda$ distribution at $z=6$. The
distribution can be approximated by a lognormal with peak $\lambda=1.07$ and
dispersion 0.28 dex. Therefore the typical quasar at $z=6$ is
observed to be accreting right at the Eddington limit and there is only a narrow
$\lambda$ distribution.

The peak of the distribution is four times higher than for the most
luminous quasars at $2<z<3$ found by Shen et al. (2008). However, the
Shen et al. work used a smaller luminosity bin than in our study, so
in order to make a detailed comparison with lower redshift we need to
use a sample that is matched in luminosity. We use the McLure \&
Dunlop (2004) \mgii\ FWHM line widths and luminosities for SDSS
quasars at $1.7<z<2.1$. We use the same equation from Vestergaard \&
Osmer (2009) as for the $z=6$ sample to determine \mbh. In order to
get luminosity-matched samples, we produce 10\,000 randomly drawn
samples of 17 quasars from the $z=2$ SDSS sample matched to the
$L_{3000}$ distribution of the $z=6$ sample. $\lambda$ is determined
for each $z=2$ quasar using the same bolometric correction as the
$z=6$ sample. The resulting $\lambda$ distribution at $z=2$ is shown
in Figure \ref{fig:eddfrac}. The distribution peaks at significantly
lower $\lambda$ than the $z=6$ sample ($\lambda=0.37$) and is somewhat
broader with dispersion 0.39 dex. A Kolmogorov-Smirnov test shows a
probability of $10^{-6}$ that these two samples are drawn from the
same parent distribution. Our results at $z=2$ are slightly different
to the results of Shen et al. (2008), likely due to the fact we have a
broader luminosity range.

The fact that virial black hole masses are calculated by $M_{\rm BH}
\propto {\rm FWHM}^2 L^{0.5}$ and that the FWHM distribution is not a
strong function of luminosity or redshift at $z<4$ has led skeptics to
suggest that the virial black hole mass estimates are purely driven by
the luminosity dependence. We have shown here a strong dependence of
FWHM with luminosity at $z=6$ and a marked difference in the $\lambda$
distribution at $z=6$ compared to $z=2$. Given the theoretical
expectation that one should see such a change at the highest redshifts
(as we discuss further below) lends strong support to the use of
virial black hole estimators at both high redshift and high luminosity
(despite the fact the method is only calibrated via the luminosity --
BLR radius relation at low redshift and low luminosity).

The observed distribution of $\lambda$ in Figure \ref{fig:eddfrac} is
that of several magnitude-limited samples of quasars (SDSS and
CFHQS). It is not therefore necessarily the same as the distribution
of $\lambda$ for a volume-limited sample of $z=6$ black holes. Black holes
with low accretion rates may not pass the quasar selection magnitude
limits. One would expect the luminosity-selected sample to have a
distribution of $\lambda$ shifted to higher values than the intrinsic
$\lambda$ distribution of all black holes with a given \mbh
\footnotemark. 

\footnotetext{Note this is different to the bias noted by Shen et
  al. (2008) for the $\lambda$ distribution in a  {\it narrow luminosity bin}
  which is biased too {\it low} due to scatter and the slope of the black hole mass function.} 

To determine the expected magnitude of this effect, we have carried
out simulations with a range of plausible black hole mass functions
and Eddington ratio distributions. In each simulation we generate
$10^6$ black holes in the mass range $3\times 10^7 - 10^{10} M_\odot$
according to the mass function and assign a Eddington ratio randomly
from a lognormal distribution. Excluding other causes of scatter (such
as scatter in the bolometric correction) this allows us to determine
the absolute magnitude of each black hole and then select absolute
magnitude limited quasar samples. Due to the steepening of the black
hole mass function at high masses (see Section \ref{bhmf}), the
absolute magnitude limit of the simulated sample is important. We
chose the results for a $M_{1450}<-25$ sample because this is the
typical magnitude of the CFHQS quasars. The observed distribution of
$\lambda$ is still a lognormal and is shifted towards higher values
from the input distribution by 0.26 dex. Therefore the intrinsic
distribution which matches our observed $z=6$ distribution has a peak
at $\lambda=0.60$. The intrinsic dispersion is only marginally broader
than that observed (0.30 dex intrinsic compared to 0.28 dex
observed). For brighter quasars, $M_{1450}<-26$, the results are very
similar with a further positive shift in $\lambda$ of 0.05 dex, i.e
about 10\%. From these simulations, we determine that the intrinsic
distribution of $\lambda$ for a volume-limited sample of $z=6$ black
holes would be a lognormal with peak $\lambda=0.6$ and dispersion 0.30
dex. This distribution, $P(\lambda)$, is plotted as a solid line on
Figure \ref{fig:eddfrac} so the relatively small offset from the
observed distribution can be appreciated.  Although there may be some
small dependence of the distribution with \mbh, there is no evidence
for a dependence in luminosity in Figure \ref{fig:bhmassfwhmlum} so we
assume no \mbh\ dependence. The observed $z=2$ Eddington ratio
distribution in Figure \ref{fig:eddfrac} would be similarly offset from the 
intrinsic $z=2$ distribution due to the same selection effect being at
work. We have not modeled this because we only need the
$z=6$ intrinsic distribution for the analysis in this paper.

\subsection{Comparison with models, lifetimes, light curves and duty cycles}
\label{modlifeduty}

In this section we compare the results above with several models for
black hole accretion growth, concentrating on the distribution of
Eddington ratios at $z=6$, implications for quasar lifetimes, light
curves and the duty cycle. If the duty cycle is close to unity,
i.e. there are no completely inactive black holes at $z=6$, then the
solid curve in Figure \ref{fig:eddfrac} shows that essentially all
$z=6$ black holes are accreting at $\lambda>0.1$ and half of them at
$\lambda>0.6$.

Di Matteo et al. (2008) used hydrodynamic cosmological simulations
to study the growth of black holes from $z=10$ to low redshift. The
accretion rate onto the black hole was governed by Bondi-Hoyle
accretion dependent upon the innermost gas density resolved by the
simulations. Feedback from the quasar becomes important in their model
when the black hole is accreting close to the Eddington limit and this
leads to quasars only spending a short amount of their lifetimes at
close to the Eddington limit (Springel et al. 2005). The $\lambda$
distribution at $z=6$ determined by Di Matteo et al. (2008) peaks at
$\lambda=0.04$ and has a negligible fraction of the distribution at
$\lambda>0.3$, which is clearly inconsistent with our distribution,
unless our black hole masses are biased low by radiation pressure
effects (Marconi et al. 2008), which we argue later is unlikely.

Sijacki et al. (2009) used a similar hydrodynamical simulation centred
on a high density peak identified in a larger dark matter
simulation. They chose a high density region in order to simulate the
growth of a massive ($\sim 10^9 M_\odot$) black hole, such as those
found by the SDSS. They adopted the same accretion and feedback
prescriptions as in Di Matteo et al. (2008). At $z=6$ they found that
all the massive ($>10^7 M_\odot$) black holes in this region were
highly accreting, with a typical $\lambda=0.5$. This is very similar
to the results we have found. Our results are for both SDSS
and CFHQS quasars and therefore likely probe a range of dark matter
densities.

A similar re-simulation of a high-density region was carried out by Li
et al. (2007). They found that during the initial growth phase at
$z>7$ of the black holes which would eventually merge by $z=6$ to form
a $>10^9 M_\odot$ black hole, the black holes accreted at
approximately the Eddington rate. As in the previous simulations, when
the black hole has grown to a certain mass at the Eddington rate where
feedback becomes important, the gas supply is shut off and the
accretion rate drops dramatically.

All the above simulations contained almost no inactive black holes at
$z=6$ (except for seed black holes which had experienced very little
growth and are below the mass range that we are interested in of $\sim
10^7 M_\odot$). Based on the plentiful gas supply in high-redshift
galaxies, the high merger rate, and Bondi-Hoyle accretion, all
black holes are expected to be active and the duty cycle is close to
unity. This is fundamentally different to the situation at lower
redshifts where many black holes have passed their peak accretion and
their host galaxies either contain gas that will not cool or have been
cleared of gas by feedback effects (e.g. Di Matteo et al. 2008).

The idea that quasar activity follows two phases; initial
Eddington-limited accretion followed by decreased or intermittent
sub-Eddington accretion has been around for a long time (Small \&
Blandford 1992).  Yu \& Lu (2008) used constraints from the local
black hole mass function and the quasar luminosity function to show
that the latter phase could be fit by quasar light curves where
luminosity declines with time with a power-law index of $-1.2$ or
$-1.3$, as would be expected if the decline phase were due to the
evolution of an isolated accretion disk (Canizzo et al. 1990). Hopkins
\& Hernquist (2009) used the luminosity function and $\lambda$
distribution at $z<1$ to place constraints on quasar light curves and
found a somewhat faster decline than the case of an isolated accretion
disk, consistent with AGN feedback effects in hydrodynamical
simulations of galaxy mergers (Hopkins et al. 2006a). They show that
quasars at low redshift spend most of their time accreting at low
Eddington rates ($\lambda \ll 0.1$). Hopkins \& Hernquist also
considered how well the derived quasar lifetimes could be translated
into quasar light curves, but found that the data are degenerate
between single long-lived accretion events taking several Gyr (of
which only the first fraction of a Gyr has high $\lambda$) or light
curves with multiple episodic short-lived accretion
events. Observations such as the transverse proximity effect (Jakobsen
et al. 2003; Worseck et al. 2007) and spectral aging studies of radio
sources (Scheuer 1995) suggest high-accretion rate episodes of length
$>10^{7}$\,yrs and therefore provide only a weak limit on the number
of potential episodes.

We have found that at $z\approx 6$ almost all moderate to high
luminosity quasars are accreting at close to the Eddington rate. The
absence of sub-Eddington quasars suggests that most quasars at this
epoch are in the process of the exponential build-up of their central
black holes and have not reached the later phase of quasar activity
where the accretion rate declines.  Only one or two of the quasars we
have studied may be in the decline part of the light curve. Taking
account of the bias in $\lambda$ due to the quasar sample magnitude
limit, this leads us to conclude that at least 50\% of black holes at
$z=6$ are still in the Eddington-limited growth phase. In addition,
the results of the simulations described above show that there are no
inactive black holes at $z=6$, i.e. the duty cycle is close to
unity. A similar conclusion was reached by Shankar et al. (2010) who
showed that the strong clustering at $z>4$ measured by Shen et
al. (2007) means quasars reside in high mass dark matter halos and
(assuming a tight $M_{BH}-M_{\rm halo}$ relation) must have a high
duty cycle of $\approx 95\%$ at $z=6$. These inferences on lifetimes
and duty cycles are important for the next section where we determine
the $z=6$ black hole mass function, because they mean that we are
directly observing most of the high mass end of the black hole mass
function at $z=6$ in quasars (with the likely exception of obscured
quasars).

\section{Black hole mass function}
\label{bhmf}

In the previous section we determined the intrinsic distribution of
Eddington ratios for supermassive black holes at $z=6$ and also
discussed their lifetimes and duty cycles. The fact that almost all
black holes at $z=6$ appear to be active and accreting at rates
approaching, or even exceeding, their Eddington rates means that it is
a relatively trivial procedure to use the observed luminosity function
of quasars to determine the underlying black hole mass
distribution. This is in contrast to the situation at lower redshifts
where a considerable fraction of black holes may have passed their
peak activity or be going through a pause in activity and therefore
one can either determine the {\it active} black hole mass
function (Vestergaard et al. 2008; Shen et al. 2008) or one requires
more model assumptions and constraints such as the Soltan argument to
estimate the total black hole mass function (Somerville 2009; Shankar
et al. 2009).

\subsection{Required inputs and assumptions}
\label{inputs}

The necessary assumptions for deriving the $z=6$ black hole mass function are:
\begin{itemize}
\item{Observed $z=6$ quasar luminosity function}
\item{Bolometric correction}
\item{Correction of luminosity function for obscured quasars}
\item{Correction for inactive black holes, i.e. duty cycle}
\item{Eddington ratio distribution}
\end{itemize}

We now discuss each of these in turn.

\subsubsection{Observed $z=6$ quasar luminosity function}

The luminosity function gives the space density of actively accreting
black holes as a function of their luminosity. The best derivation of
the luminosity function at $z=6$ (Willott et al. 2010) comes from
optically-selected quasars from the SDSS and CFHQS surveys. Ideally,
hard X-ray or mid-IR samples of quasars would be used since they
account for some fraction of the optically-obscured
population. However, there are no AGN selected at X-ray or mid-IR
wavelengths at this high a redshift. 

\begin{figure}
\hspace{-0.2cm}
\resizebox{0.50\textwidth}{!}{\includegraphics{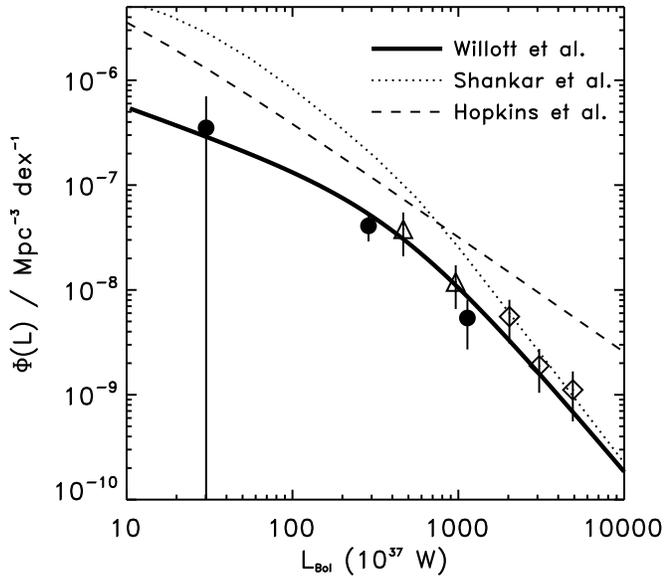}}
\caption{Comparison of bolometric quasar luminosity functions at $z=6$ corrected for the existence of obscured quasars as described in Section \ref{inputs}. The binned data points and thick black line are from the optical quasar luminosity function of Willott et al. (2010) using data from the SDSS main (diamonds -- Fan et al. 2006), SDSS deep stripe (triangles -- Jiang et al. 2009) and CFHQS (filled circles). Error bars on the binned data are only the Poisson errors on the quasar counts and do not include uncertainties in the bolometric correction or obscured AGN fraction. The dotted line is the luminosity function evolutionary model of Shankar et al. (2009) and the dashed line that of Hopkins et al. (2007). The CFHQS and SDSS deep stripe data are inconsistent with both model luminosity functions.
\label{fig:bollf}
}
\end{figure}

The luminosity function derived by Willott et al. (2010) used 40
quasars at $5.74<z<6.42$. These comprised 14 quasars from the SDSS
main sample (Fan et al. 2006), 10 quasars from the SDSS deep stripe
(Jiang et al. 2009) and 16 quasars from the CFHQS (Willott et
al. 2010) giving a broad range in luminosity. The luminosity function
was fit by a double power law model with best fit parameters of
$\Phi(M_{1450}^{*})= 1.14\times 10^{-8}\,{\rm Mpc}^{-3}\,{\rm
  mag}^{-1}$, break magnitude $M_{1450}^{*}=-25.13$ and bright end
slope $\beta=-2.81$. A faint end slope of $\alpha=-1.5$ was
assumed. Although there is considerable uncertainty on the values of
these parameters, due to covariance of the parameters, the space
density of optical quasars at $z=6$ is strongly constrained
(1\,$\sigma$ uncertainty $<0.1$\,dex) over the range
$-27.5<M_{1450}<-24.7$. The faintest quasar used, CFHQS\,J0216-0455 at
$z=6.01$, has absolute magnitude $M_{1450}=-22.2$, but this being
the only quasar at $M_{1450}>-24$ means the luminosity function is
quite uncertain at low luminosities.

\subsubsection{Bolometric correction}

As in Section \ref{bhdeterm} we use a bolometric correction from the
work of Richards et al. (2006) to convert between monochromatic UV
luminosity and bolometric luminosity. Because the quasar luminosity
function is defined at 1450\,\AA\ we use a luminosity correction
factor of 4.4. The bolometric correction is not an important factor
here because the black hole mass is directly determined from the
virial mass estimator which uses the observed UV continuum
luminosity. We therefore do not include any extra scatter due to the
bolometric correction.

\subsubsection{Correction of luminosity function for obscured quasars}

The optical quasar luminosity function does not account for the
population of actively accreting black holes which have their optical
(in this case rest-frame far-UV) radiation absorbed by dust. In order
to determine the full population of accreting black holes, one needs
to correct for the missing population. The obscured AGN may be
obscured by a geometrical torus as required by the simple unified
scheme (Antonucci 1993; Urry \& Padovani 1995) or by dust in the
quasar host galaxy (Martinez-Sansigre et al. 2005). It is now well
established that low luminosity AGN are more frequently obscured than
those of higher luminosity (Lawrence 1991; Ueda et al. 2003) and
therefore we need to adopt a correction that is
luminosity-dependent. Unfortunately, there is no evidence yet on the
redshift dependence of the obscured fraction up to $z=6$. There
appears to be no evolution in this fraction from $z=0$ to $z=2$ (Ueda
et al. 2003). Therefore we take the low redshift absorbed AGN fraction
as a function of luminosity as determined by Ueda et al. (2003). We
also include Compton thick AGN following Shankar et al. (2009),
although this only makes a modest further increase. Including obscured
quasars raises the space density by a factor of 2 at $M_{1450}=-27.2$
and a factor of 3 at $M_{1450}=-20.7$.

Figure \ref{fig:bollf} shows the bolometric luminosity function of
Willott et al. (2010) using the bolometric correction and correction
for obscured quasars described above. Also plotted are $z=6$
bolometric luminosity function models from Hopkins et al. (2007) and
Shankar et al. (2009; based on the compilation of Shankar \& Mathur
2007). Both the Shankar et al. and Hopkins et al. luminosity
functions over-predicts the numbers of quasars compared to the
observations. This is discussed further in Section \ref{derivbhmf}.

\subsubsection{Correction for inactive black holes, i.e. duty cycle}

As discussed at the end of Section \ref{modlifeduty}, there are both
theoretical and observational arguments that, at $z=6$, the duty cycle
of black holes is very high. We adopt a default duty cycle of 0.75 and
consider a plausible range of 0.5 to 1.

\subsubsection{Eddington ratio distribution}

Using a constant bolometric correction means that a given black hole
mass maps directly on to a value of $M_{1450}$ for black holes
accreting at exactly the Eddington limit. The Eddington ratio
distribution, $P(\lambda)$, is used to map a given black hole mass
onto a range of $M_{1450}$. We use the {\it intrinsic} Eddington ratio
distribution at $z=6$ determined in Section \ref{bh} and plotted in
Figure \ref{fig:eddfrac} as a solid line which is a lognormal centred
at $\lambda=0.6$ with standard deviation of 0.30 dex. We do not apply
a cutoff above the Eddington limit, as quasars are observed to have
values up to $\lambda=10$ (although of course some of this is due to
scatter in the measurements and correlations used to determine black
hole masses). In any case, this distribution only puts 4\% of black
holes at $\lambda>2$. We do not have sufficient data to determine if
there is any black hole mass dependence of $P(\lambda)$. At the lowest
luminosities and black hole masses, we do not have any quasars with
black hole masses from the \mgii\ line so we assume that $P(\lambda)$
is the same at all luminosities. We note that the lowest luminosity
quasar in the CFHQS sample, CFHQS\,J0216-0455, does have a very narrow
\lya\ line which would have an intrinsic FWHM of $1600\,{\rm
  km\,s}^{-1}$ correcting for blue wing IGM absorption (Willott et
al. 2009). If \mgii\ has a similar width, this quasar would be
accreting at the Eddington limit. We assumed the population of
obscured AGN has the same Eddington ratio distribution as that of the
unobscured population.

\subsubsection{Deriving the black hole mass function}
\label{derivbhmf}

Rather than attempt to invert the observed luminosity function of
Willott et al. (2010) to determine the black hole mass function, we
instead use model black hole mass functions to produce luminosity
functions which are then fit to the same quasar samples as Willott et
al. (2010). Therefore much of the fitting procedure is the same as the
luminosity function fitting and we refer the interested reader to that
paper for full details. The best-fit is determined via the maximum
likelihood method using amoeba parameter optimization.

The main effect of including scatter in the conversion of black hole
mass to quasar luminosity is to flatten the bright end slope of the
luminosity function compared to the slope for the black hole mass
function. Fits to the black hole mass function were attempted using
either double power laws or Schechter functions. The double power law
fits produced a very steep high mass end and similar
likelihood to the Schechter function fits. Therefore we adopted the
Schechter function for the black hole mass function since it required
one fewer parameter to be fit. The Schechter function form is favoured
theoretically because there should be a sharp cutoff in mass of
the most massive black holes, as is observed in the local black hole mass
distribution (e.g. Shankar et al. 2009) and the $z=2$ black hole mass distribution
(Vestergaard et al. 2008), which is plausibly due to feedback in the
most massive galaxies, and the limited cosmic time available for black hole
growth by $z=6$ (Volonteri \& Rees 2005).

The process then is to take the model black hole mass function,
convolve with $P(\lambda)$, convert to luminosity and absolute
magnitude using the bolometric correction, correct for obscured AGN
and the duty cycle to generate a model luminosity function. This model
luminosity function is compared to the data and the parameters of the
black hole mass function optimized to generate the best fit that
maximizes the likelihood. The $z=6$ black hole mass function,
$\Phi(M_{\rm BH})$, assuming a duty cycle of 0.75, has best fit
parameters of $\Phi(M_{\rm BH})^{*}= 1.23\times 10^{-8}\,{\rm
  Mpc}^{-3}\,{\rm dex}^{-1}$, characteristic mass $M_{\rm
  BH}^{*}=2.24\times 10^{9} M_\odot$ and faint end slope
$\alpha=-1.03$.

\begin{figure}
\hspace{-0.2cm}
\resizebox{0.50\textwidth}{!}{\includegraphics{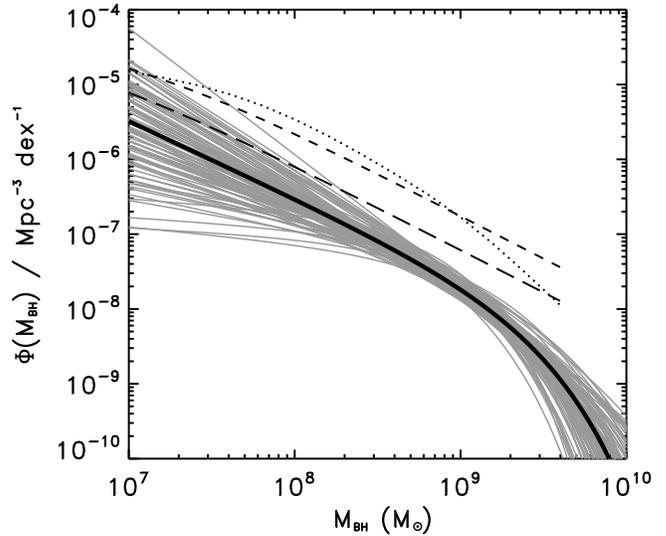}}
\caption{$z=6$ black hole mass function determined in Section \ref{bhmf}. The thick solid black line is the best-fit to the data. The thin gray lines are the 100 bootstrap resamples consistent with the data. The black hole mass function is most strongly constrained at $M_{\rm BH} >  10^{8} M_\odot$ and at $M_{\rm BH}<3\times 10^{9} M_\odot$. Also plotted are $z=6$  black hole mass functions determined from the evolutionary models of Shankar et al. (2009) assuming their luminosity function and reference model ($\epsilon=0.065$, $\lambda=0.4$, $z=6~ {\rm duty~ cycle}=0.5$) (dotted line) and assuming the Hopkins et al. (2007) luminosity function with the reference model (short-dashed line) and also with different parameters of $\epsilon=0.09$, $\lambda=1$, $z=6 ~ {\rm duty~ cycle}=0.5$ (long-dashed line).
\label{fig:mf}
}
\end{figure}

To determine the plausible range of black hole mass functions
consistent with the data, we employ bootstrap resampling as described
in Willott et al. (2010). We generate 100 samples of the data and
determine the best fit to each. To account for the uncertainty in the
duty cycle at $z=6$ we take the duty cycle as a uniformly distributed
random number between 0.5 and 1 for each resample. In this way, we
generate 100 plausible black hole mass functions.

Figure \ref{fig:mf} shows the 100 bootstrap resampled mass functions
as overlapping gray lines and the best fit as a thick black line. The
mass function is most strongly constrained close to $10^{9}
M_\odot$. This is not surprising given the range of black hole masses
of SDSS and CFHQS quasars shown in Figure
\ref{fig:bhmassfwhmlum}. There is considerable divergence at $M_{\rm
  BH} < 10^{8} M_\odot$ and at $M_{\rm BH}>3\times 10^{9} M_\odot$ due
to the few very low or very high luminosity quasars used to derive the
luminosity function.

Also shown on Figure \ref{fig:mf} are three determinations of the
$z=6$ black hole mass function by Shankar et al. (2009). These were
generated by self-consistent models which fit the evolving AGN
luminosity function, local black hole mass function and X-ray
background and are taken from their Table 3. One curve is for the
Shankar et al. reference model and the other two are using the same
model but fitting the Hopkins et al. (2007) luminosity function rather
than that compiled by Shankar et al. The black hole mass function
determined by us is considerably lower than all three of the Shankar
et al. models, by factors ranging from $3$ to $10$. We will next
investigate the reasons for these differences.

Most cosmic accretion occurs at $z\ll 6$ and therefore the local black
hole mass function and cosmic X-ray background are not strong
constraints on the $z=6$ black hole mass functions derived by Shankar
et al. (2009).  The main reason for the difference between our result
and theirs comes from their luminosity functions (Figure
\ref{fig:bollf}) and parameters used to convert between luminosity and
black hole mass. The luminosity function derived by Shankar et
al. (2009) agrees well with the bright end of the $z=6$ luminosity
function of Fan et al. (2004) but is significantly steeper than that
determined by Willott et al. (2010). Given that Willott et al. (2010)
also found a slightly lower bright end normalization, means that our
best-fit luminosity function is 3 times lower than the reference model
of Shankar et al. (averaged over $-28<M<-23$). The other luminosity
function used by Shankar et al. is that of Hopkins et al. (2007). As
seen in Figure \ref{fig:bollf} this function gives fewer moderate
luminosity quasars than the Shankar et al. luminosity function, but
more high luminosity ones. This can be seen also for the black hole
mass functions in Figure \ref{fig:mf}. The Hopkins et al. luminosity
function is on average 4 times higher than that of Willott et
al. averaged over $-28<M<-23$. At moderate quasar luminosities
($-26<M<-24$) the space densities of both the Shankar et al. and
Hopkins et al. luminosity functions are strongly ruled out by the data
of Willott et al.  Therefore differing luminosity functions account
for a large part of the difference between our black hole mass
function and that of Shankar et al.

The rest of the difference can be accounted for by the parameters used
to translate black hole masses to luminosities and the duty cycle. The
relevant parameters for the reference model of Shankar et al. (2009)
are accretion efficiency $\epsilon=0.065$, constant Eddington ratio
$\lambda=0.4$ and duty cycle at $z=6$ of 0.5. Only the duty cycle
evolves with redshift in the reference model. In comparison, our
analysis assumes $\epsilon=0.09$, a distribution of $\lambda$ peaked
at 0.6 and a duty cycle drawn from a uniform distribution between 0.5
and 1 for the bootstrap resamples. The long-dashed line shows the model using the
Hopkins et al. (2007) luminosity function with different parameters
($\epsilon=0.09$, $\lambda=1$), which were required to fit the local
black hole mass function. One can see that the effect of using these
parameters, which are similar to ours, shifts the black hole mass
function lower by a factor of $\sim 3$. Therefore we see that the
differences in duty cycle, accretion efficiency, Eddington ratio and
luminosity functions cause the differences between our results and
those of Shankar et al. (2009). Both the Eddington ratio and
luminosity function have been determined by us and therefore are an
improvement on previous work. The duty cycle and accretion efficiency
are still very much unknown.

The hydrodynamic simulations of Di Matteo et al. (2008) make
predictions for the evolution of the black hole mass function. Due to
the relatively small simulation box size, their results are only valid
for black holes up to $M_{\rm BH} \sim 10^{8} M_\odot$. They predict a
space density of $z=6$ black holes a factor of $\sim 100$ greater than
we do. In large part this is due to their very much lower assumed
accretion rates discussed in Section \ref{modlifeduty}. Our quasars are
fewer in number but growing much more rapidly than those in their
simulations.

Marconi et al. (2008) derived corrections to virial black hole mass
estimates based on the effect that radiation pressure has on the gas
velocities. They showed that close to the Eddington limit this could
have a very large effect with black hole masses being underestimated by
a factor of $\sim 10$ assuming the BLR cloud column density is $N_{\rm
  H}=10^{23}{\rm cm ^{-2}}$. We did not use the Marconi et
al. corrections due to the unknown value of $N_{\rm H}$. We note that
because the $z=6$ quasars appear to be accreting close to the
Eddington limit, this correction would be so large that it would shift
the entire black hole mass function to higher masses by 1 dex. This
would then give the exponential cutoff in the mass function at $M_{\rm
  BH} \sim 3 \times 10^{10} M_\odot$. It does not seem plausible that
the mass function at $z=6$ would contain such a large number of $>
10^{10} M_\odot$ black holes given the lack of such black holes at low
redshift (Tundo et al. 2007), at $z=2$ (McLure \& Dunlop 2004;
Vestergaard \& Osmer 2009), self-regulation arguments (Natarajan \&
Treister 2009) and the difficulty of forming them in the cosmological
time available by $z=6$ (Volonteri \& Rees 2006). The issue of whether
or not one should make radiation pressure modifications to virial
black hole mass estimates is also discussed by Peterson (2010).

\subsection{Global evolution of black hole and stellar mass functions}

\begin{figure}
\hspace{-0.2cm}
\resizebox{0.50\textwidth}{!}{\includegraphics{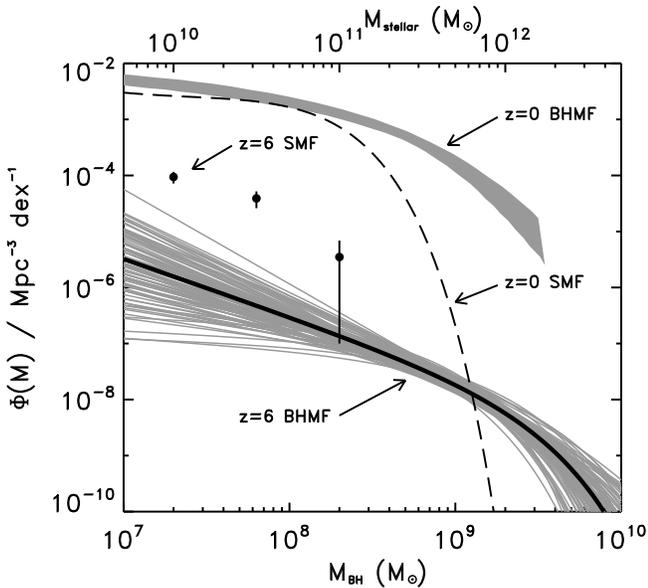}}
\caption{The black hole mass function at $z=6$ ($z=6$ BHMF; thick black line and
  grey curves as in Figure \ref{fig:mf}) compared with the local black
  hole mass function ($z=0$ BHMF; grey band -- Shankar et al. 2009), the local
  stellar mass function ($z=0$ SMF; dashed line -- Baldry et al. 2008) and the
  $z=6$ stellar mass function ($z=6$ SMF; circles with error bars -- Stark et
  al. 2009). The two stellar mass functions have been shifted horizontally by a
  factor of $2\times 10^{-3}$, as indicated in the upper axis, to
  match up the space densities of galaxies and black holes at $z=0$.
\label{fig:mfz0z6}
}
\end{figure}

The $z=6$ black hole mass function we have derived can be compared to
the present day black hole mass function in order to determine the
total black hole growth between $\sim 1$\,Gyr after the Big Bang to
today. Locally, most black holes are inactive and measurement of their
masses via dynamical tracers such as gas and stars have only been
performed for about 50 of the closest galaxies (Gultekin et
al. 2009). Given the small scatter in scaling relations such as the
$M_{\rm BH}-\sigma$ and $M_{\rm BH}-L$ relations, it is common to
derive the local black hole mass function by using the observed
velocity dispersion function or luminosity function of galaxies in
conjunction with scaling relations (see Tundo et al. 2007 for a
comparison of these methods). We use the range of plausible local
black hole mass functions determined from analysis by Shankar et
al. (2009) of several such studies (plotted as a gray band in Figure
\ref{fig:mfz0z6}).

The local and $z=6$ black hole mass functions shown in Figure
\ref{fig:mfz0z6} have rather similar shapes. The local function
steepens above a somewhat lower mass of $\sim 10^9 M_\odot$ than the
$z=6$ function. There is also some evidence for a steeper $z=6$
function at $10^7 M_\odot < M_{\rm BH} < 3 \times 10^8 M_\odot$,
although the $z=6$ function is not well constrained at the low mass
end, due to few low-luminosity quasars and lack of $\lambda$
measurements for them.  The overall normalization of the $z=6$ black
hole mass function is $\sim 10^{-4}$ times that at $z=0$, highlighting just
how rare black holes are at this early epoch.

Due to the existence of local scaling relations between black holes
and galaxy properties and the implications for the importance of black
holes in galaxy evolution, it is interesting to compare this evolution
of the black hole mass function with the evolution of the mass which
has formed into stars within galaxies. For the local stellar mass
function we use the $z<0.05$ determination using the SDSS New York
University Value-Added Galaxy Catalog (Baldry et al. 2008). For the
stellar mass function at $z=6$ we use that determined via star
formation history SED fits to $z\approx 6$ Lyman break galaxies with
deep {\it Spitzer} photometry by Stark et al. (2009). These data start
to become incomplete at $M_{\rm stellar} < 10^{10} M_\odot$ and have a
large uncertainty at $M_{\rm stellar} = 10^{11} M_\odot$ due to the
small volumes probed by these Lyman break surveys.

In Figure \ref{fig:mfz0z6} we plot the $z=0$ and $z=6$ stellar mass
functions along with the black hole mass functions. The upper axis
shows the stellar mass and it has been shifted horizontally by $2\times 10^{-3}$
compared to the black hole mass axis. This factor is similar to the factors of
$1.4\times 10^{-3}$ (Haring \& Rix 2004) and $2\times 10^{-3}$ (Tundo
et al. 2007) determined from fits to the local $M_{\rm BH}-M_{\rm
  bulge}$ relation and approximately matches the space densities of
galaxies and black holes at $z=0$ in Figure \ref{fig:mfz0z6}.

Figure \ref{fig:mfz0z6} shows the rather remarkable result that
whereas the stellar mass density evolution from $z=6$ to $z=0$ is
$\sim 100$, the black hole mass density evolution over this same time
period is $\sim 10^{4}$, i.e. 100 times greater. This is observed to
hold over about one order of magnitude in mass. At low black hole
masses and high stellar masses the comparison is difficult because of
the vastly different volume surveys undertaken for quasars and
galaxies at $z=6$. At $M_{\rm stellar} = 10^{11} M_\odot$, the very
large uncertainty in stellar mass density makes it plausible that it
matches that of black holes. Larger area optical to mid-IR surveys for
$\sim 10^{11} M_\odot$ galaxies at $z\approx 6$ are
needed to address whether there is an exponential decline in the
stellar mass function at this mass which would match the space density
of $M_{\rm BH} = 2 \times 10^8 M_\odot$ black holes.

We identify two possible reasons for this big difference between the
mass accreted on to black holes and that which has formed stars by
$z=6$. Eddington-limited accretion occurs exponentially so at early
times when the black hole masses are small, there is a limit to how
fast they can grow. In comparison, star formation has no similar
restriction and in fact the high densities in the early universe may
actually have enhanced the star formation rate (e.g. Granato et
al. 2004). Another possibility is that only a fraction of galaxies
were seeded with massive $\gtsimeq 10^{4} M_\odot$ black holes. Most
theoretical attempts to account for the most massive black holes
observed at $z=6$ have shown that seed black holes of this mass are
required if accretion obeys the Eddington limit (Yoo \&
Miralda-Escud\'e 2004; Li et al. 2007; Sijacki et al. 2009). Therefore
the finding of a large discrepancy between the space densities of
black holes and galaxies may be indicating that only a fraction of
galaxies start with massive seeds. The less massive seeds in other
galaxies will catch up by the peak in quasar activity at $z\sim2$.

This result is in contrast to observations of the evolution of the
$M_{\rm BH}-M_{\rm bulge}$ relation. In several studies it has been
found that for AGN-selected samples, the ratio of $M_{\rm BH}$ to
$M_{\rm bulge}$ evolves to higher values at higher redshift (Walter et
al. 2004; Peng et al. 2006; McLure et al. 2006; Riechers et al. 2008;
Merloni et al. 2010; Wang et al. 2010). It is important to note that
due to scatter in the $M_{\rm BH}$ to $M_{\rm bulge}$ (or $M_{\rm
  halo}$) relation combined with steep mass functions, luminous
quasars will preferentially be found in lower mass hosts than would be
expected based on the relation for a volume-limited sample of galaxies
(Willott et al. 2005; Lauer et al. 2007). Somerville (2009) and
Merloni et al. (2010) show that this bias could almost completely
account for the observed evolution of the relation and that up to
$z=2$ there is at most a factor of two of positive evolution in
$M_{\rm BH}/M_{\rm bulge}$. Other studies, selecting galaxies to be
gas-rich starbursts without luminous AGN (Borys et al. 2005; Alexander
et al. 2008), actually show negative evolution in $M_{\rm BH}/M_{\rm
  bulge}$ highlighting how important selection effects are for
distinct galaxy populations. Note that our results are not necessarily
in conflict with the molecular gas observations of $z\sim 6$ quasars
that show they have high ratios of $M_{\rm BH}$ to $M_{\rm bulge}$
(Walter et al. 2004; Riechers et al. 2008; Wang et al. 2010). The
point is that there are many more galaxies with very low ratios of
$M_{\rm BH}$ to $M_{\rm bulge}$ which are not observed in quasar
samples.

Attempts to study the theoretical global evolution in star formation
versus black hole growth have been carried out by Robertson et
al. (2006), Hopkins et al. (2006b) and Shankar et al. (2009). These
show little evolution in the ratio, however they mostly do not go up
to $z=6$ or are poorly constrained there. Di Matteo et al. (2008)
consider the evolution of the stellar mass density and black hole mass
density in their simulations. They find that although there is little
evolution from the local ratio up to $z=2$, there is considerable
evolution from $z=2$ to $z=6$ of nearly a factor of 10 (in the sense
that the black hole mass density declines more rapidly). Their results
are similar to ours, but with less rapid evolution, which can be
explained because their simulations have a lower typical $\lambda$
than we observe and therefore a higher black hole mass function. Li et
al. (2007) found in their simulations that the total stellar mass
formed in the galaxies which would merge to become a $z=6$ SDSS host
galaxy led the total black hole mass by factors of 10 to 100 at early
times ($8<z<14$). Although not discussed in their paper, the likely
reason for this behaviour is the Eddington limit which restricts the
very early growth of the black hole mass.  Lamastra et al. (2010) use
a semi-analytic cosmological model to predict positive evolution of
$M_{\rm BH}/M_{\rm bulge}$ by a factor of 3 for the global population
up to $z=7$ which they explain as due to very efficient black hole
growth at early times. Their work suggests a ratio of black hole to
stellar mass density $\sim 300$ times greater than we have found!

Our results hinge upon the assertions that the quasar duty cycle is
very high at $z=6$, there is little evolution in the fraction of
obscured AGN and therefore we have identified all the black holes as
AGN. The $z=6$ Lyman break galaxies with stellar masses
$>10^{10}M_\odot$ are about 100 times more common than low-luminosity
quasars. It is possible that these galaxies contain inactive or
obscured black holes and we have underestimated the true space density
of black holes. We did account for a factor of 2--3 of obscured
quasars, as observed at lower redshift, but perhaps this factor
evolves dramatically at the highest redshifts due to the dusty nature
of these young, forming host galaxies.  Mid-IR-selected AGN samples do
show an increase in the obscured fraction at $z>3$ (M. Lacy,
priv. comm.). Although there are observations requiring large dust
masses in many $z\sim 6$ quasar host galaxies (Wang et al. 2008), the
Lyman break-selected $z\sim 6$ galaxies have very low dust extinction
based on their UV spectral slopes (Bouwens et al. 2009).

At $z\sim 2 - 3$, about 3--5\% of Lyman break galaxies contain
evidence for weak AGN (Steidel et al. 2002; Reddy et
al. 2006). Spectra of $z=6$ galaxies show a high incidence of narrow
\lya\ emission lines (Stanway et al. 2007; Ouchi et a. 2008), but most
higher ionization metal lines which could be used to test for AGN lie
in the near-IR. The NIRSpec instrument on the {\it James Webb Space
  Telescope} will be able to detect weak high ionization lines and/or
weak broad Balmer lines in these galaxies to determine if they do host hidden
AGN. Another potential method of searching for obscured AGN activity
is mid-IR imaging of the hot dust with the MIRI instrument.

\section{Conclusions}

We have presented new near-IR spectroscopy for nine of the most
distant known, moderate luminosity quasars. These data have been used
to estimate virial black hole masses based on the \mgii\ linewidth and
UV luminosity. Adding in published data on more luminous quasars we
obtain a sample of 17 quasars at $z \gtsimeq 6$. Our results and
conclusions can be summarized as follows.

\begin{itemize}

\item{We observe a positive correlation between the \mgii\ line FWHM
  and UV continuum luminosity. Such a correlation is usually absent
  from lower redshift samples due to the broad range of Eddington
  ratios at lower redshift (Fine et al. 2008; Shen et al. 2008). The
  existence of this correlation at high-redshift provides support for
  the validity of the virial linewidth estimator.}

\item{There is a linear correlation between black hole mass and UV
  continuum luminosity and the quasars are accreting close to the
  Eddington limit. The distribution of {\it observed} Eddington ratios
  is a lognormal centred on $\lambda=1.07$ with dispersion 0.28
  dex. This distribution is significantly different from that of a
  luminosity-matched $z=2$ sample which has a distribution centred on
  $\lambda=0.37$ with a broader dispersion of 0.39 dex. Accounting for
  selection effects due to the quasar sample magnitude limits, we
  determine the {\it intrinsic} Eddington ratio distribution for a
  volume-limited sample of black holes at $z=6$. This distribution is
  a lognormal centred on $\lambda=0.60$ with dispersion 0.30 dex. The
  Eddington ratio distribution we find at $z=6$ is consistent with the
  results of simulations of high-density peaks in the early dark
  matter distribution (Li et al. 2007; Sijacki et a. 2009).}

\item{The implication of these results is that at $z=6$ all the
  quasars we are observing are still in their initial exponential
  growth phases due to their young host galaxies and a plentiful gas
  supply. Combining the Eddington ratio distribution with the
  assumption of a high duty cycle and the observed quasar luminosity
  function of Willott et al. (2010) we can derive the black hole mass
  function. Note that this is a much harder problem at low redshifts
  where the Eddington ratio distribution is broader and the duty cycle
  is unknown. The resulting black hole mass function is factors of $3$
  to $10$ below previous estimates (Shankar et al. 2009) due to a
  lower luminosity function normalization, accretion radiative efficiency, duty cycle and higher Eddington
  ratios.}

\item{The evolution in the black hole mass function from $z=6$ to
  $z=0$ is a factor of $\sim 10^4$. This is much greater than the
  $\sim 10^2$ increase in the stellar mass function over the same
  redshift interval. This means that the stars in galaxies were formed
  much more rapidly at high redshift than black holes grew, presumably
  due either to the limited rate at which black holes can grow due to
  radiation pressure or that only a small fraction of galaxies had
  massive initial black hole seeds.}

\end{itemize}

\acknowledgments

Thanks to Ross McLure, Francesco Shankar and Dan Stark for providing
data in electronic form and the anonymous referee for useful
suggestions for improvements. Based on observations obtained with
MegaPrime/MegaCam, a joint project of CFHT and CEA/DAPNIA, at the
Canada-France-Hawaii Telescope (CFHT) which is operated by the
National Research Council (NRC) of Canada, the Institut National des
Sciences de l'Univers of the Centre National de la Recherche
Scientifique (CNRS) of France, and the University of Hawaii. This work
is based in part on data products produced at TERAPIX and the Canadian
Astronomy Data Centre as part of the Canada-France-Hawaii Telescope
Legacy Survey, a collaborative project of NRC and CNRS. Based on
observations obtained at the Gemini Observatory, which is operated by
the Association of Universities for Research in Astronomy, Inc., under
a cooperative agreement with the NSF on behalf of the Gemini
partnership: the National Science Foundation (United States), the
Particle Physics and Astronomy Research Council (United Kingdom), the
National Research Council (Canada), CONICYT (Chile), the Australian
Research Council (Australia), CNPq (Brazil) and CONICET (Argentina).
This paper uses data from Gemini programs GS-2006A-Q-16,
GS-2009B-Q-25, GN-2007A-Q-201, GN-2007B-Q-35, GN-2008B-Q-43,
GN-2009B-Q-33 and GN-2009B-DD-5. Based on observations made with the
ESO New Technology Telescope at the La Silla Observatory.

\clearpage

\appendix

\section{Spectra which did not yield useful estimates on black hole masses}

Four CFHQS quasars were observed with the NIRI spectrograph and did
not yield measurements of the \mgii\ emission line width. In all but
one case, this was due to insufficient S/N for these very faint
targets. In the other case it is because the quasar does not have
broad emission lines.  Therefore black hole masses could not be
determined. Here we show the spectra and discuss why \mgii\ lines
could not be measured and what constraints exist on the broad line widths
in these quasars. None of these quasars are likely to have broad lines
significantly broader than the quasars in Section \ref{nirspec} and therefore the
exclusion of these four quasars does not significantly bias the
results and analysis presented in this paper.

\begin{figure*}
\resizebox{1.00\textwidth}{!}{\includegraphics{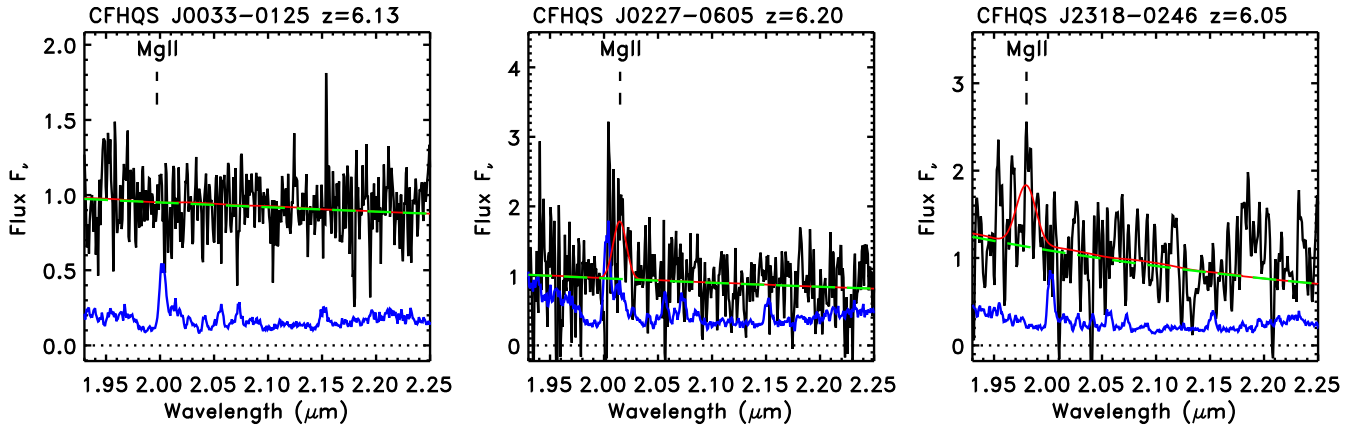}}
\vspace{-0.4cm}
\caption{$K$-band spectra of three of the four CFHQS quasars (black line) which did not yield a useful measurement of the \mgii\ line width. Best-fit model of power law continuum, broadened Fe template and broadened
  \mgii\ doublet is shown as a red line. The power law continuum only
  is shown as a green dashed line. At the bottom of the plot, the blue
  line is the $1\sigma$ noise spectrum. The expected location of the
  \mgii\ broad emission line is labeled, based on the published \lya\ redshift.
\label{fig:badnirispec}
}
\end{figure*}
\vspace{-0.5cm}

\subsection{CFHQS\,J0033-0125}

The optical spectrum of this quasar is very unusual because it shows
an extremely weak \lya\ line at $z=6.13$ (Willott et
al. 2007). Unpublished higher resolution spectroscopy confirms the
absence of significant \lya\ emission, but does show a sharp continuum
break at a \lya\ redshift of $z=6.10$ and intervening metal absorption
at redshifts up to $z=6.0$ confirming the high-redshift nature of this
quasar. The near-IR spectrum has relatively high continuum S/N in the
range $5 - 10$ at the expected \mgii\ wavelength. As can be seen in
Figure \ref{fig:badnirispec} there is absolutely no sign of an
\mgii\ line in this quasar, so the lack of \lya\ is not just due to
dust or IGM absorption, but a lack of broad emission lines such as
the quasars investigated by Diamond-Stanic et al. (2009).

\subsection{CFHQS\,J0216-0455}

This quasar is the faintest $z\sim 6$ quasar known by some margin with
an absolute magnitude $M_{1450}=-22.2$ and could be considered a
Seyfert galaxy rather than quasar based on luminosity (Willott et
al. 2009). Due to its faintness in the near-IR ($K_{\rm AB} \approx
24$) we did not expect high S/N continuum in the NIRI spectrum (which
was only composed of 1.75 hours of good weather data). The only reason
that we targeted this extremely faint quasar was because the
\lya\ line is very strong and if the \mgii\ line were comparably
strong, it could have been possible to measure it in $\approx 4$ hours
on-source. Neither continuum nor a \mgii\ line could be seen in the
1.75 hours NIRI spectrum and therefore no spectrum was extracted. We
note that the \lya\ line is very narrow with FWHM of $1600\,{\rm
  km\,s}^{-1}$ (after correcting for absorption of the blue wing) which
implies it has a very high accretion rate, close to the Eddington
limit.

\subsection{CFHQS\,J0227-0605}

The quasar has $z=6.20$ based on the \lya\ line position (Willott et
al. 2009). It is quite faint ($M_{1450}=-25.03$) and the
continuum S/N per pixel in the 2.6 hour long NIRI integration is only
$\sim 3$. As shown in Figure \ref{fig:badnirispec} the \mgii\ emission
line is detected, but at this redshift, some parts of the line
co-incide with the noisiest regions of our spectra due to a
combination of a strong sky line and atmospheric absorption. This
prevents a reliable measurement of either the line center or the
FWHM. The \mgii\ line appears fairly narrow and the best fit has $z=6.20$ and FWHM\,=\,$1900 ~{\rm km\,s}^{-1}$.
However, we do not include it in our analysis because the FWHM is very uncertain.

\subsection{CFHQS\,J2318-0246}

This is another very faint quasar ($M_{1450}=-24.83$) with a
\lya\ line at $z=6.05$ (Willott et al. 2009). The 3.9 hours NIRI
spectrum shows several peaks around the expected \mgii\ wavelength
(Figure \ref{fig:badnirispec}). Due to the low S/N it is difficult to
constrain the FWHM. The best fit line is at $z=6.07$ and has
FWHM\,=\,$3100 ~{\rm km\,s}^{-1}$. Note that the best fit has a very
steep, blue continuum which, if extrapolated, is not consistent with the
known $z'$ and $J$ magnitudes.

\end{document}